# Is It Always Windy Somewhere? Occurrence of Low-Wind-Power Events over Large Areas

Mark A. Handschy,[1,2] Stephen Rose,[3] and Jay Apt[3,4]

**Author Affiliations**
[1] Enduring Energy, LLC, 5589 Arapahoe Ave., Suite 203, Boulder, Colorado, USA
[2] Cooperative Institute for Research in the Environmental Sciences, University of Colorado, Boulder, Colorado, USA
[3] Department of Engineering & Public Policy, Carnegie Mellon University, Pittsburgh, USA
[4] Tepper School of Business, Carnegie Mellon University, Pittsburgh, USA

**Corresponding Author**
Mark A. Handschy
markh@enduringenergyllc.com

**Abstract**
The incidence of widespread low-wind conditions is important to the reliability and economics of electric grids with large amounts of wind power. In order to investigate a future in which wind plants are geographically widespread but interconnected, we examine how frequently low generation levels occur for wind power aggregated from distant, weakly-correlated wind generators. We simulate the wind power using anemometer data from nine tall-tower sites spanning the contiguous United States. We find that the number of low-power hours per year declines exponentially with the number of sites being aggregated. Hours with power levels below 5% of total capacity, for example, drop by a factor of about 60, from 2140 h/y for the median single site to 36 h/y for the generation aggregated from all nine sites; the standard deviations drops by a factor of 3. The systematic dependence of generation-level probability distribution "tails" on both number and power threshold is well described by the theory of Large Deviations. Combining this theory for tail behavior with the normal distribution for behavior near the mean allows us to estimate, without the use of any adjustable parameters, the entire generation duration curve as a function of the number of essentially independent sites in the array.

**Introduction**

The benefit of geographic diversity in reducing the variability of wind power has been investigated since integration of wind generation into the electric grid was first seriously considered. Based on his analysis of 5,000 wind speed data points recorded by the U.S. Weather Bureau at twenty cities east of the Mississippi River, Thomas speculated in 1945 that firm capacities of 50–60% of average generation could be obtained [1], while shortly thereafter Putnam assessed the capacity value of geographic diversity to be worth less than the cost of transmission needed to achieve it [2]. The smoothing benefit provided by geographic diversity would have considerable economic importance if it allowed a grid system to meet reliability targets with less conventional "dispatchable" generating capacity than would otherwise be needed for a similar amount of unsmoothed wind power. In the terminology of grid reliability it is useful to ask to what extent geographic diversity increases wind power's firm generating capacity or its effective load carrying capacity (ELCC) [3-6].

The probability that the aggregated power from an array of wind generators falls below some small generation level is of particular importance in determining ELCC and reserve requirements, as pointed out by Kahn [3] and by Haslett and Diesendorf [4]. Characterizing such "tail" probabilities and modeling how they depend on factors such as the number and geographic layout of wind plants making up the array can be challenging. Conventional measures of variation around the mean, such as the variance or standard deviation, reveal little about tail probabilities. Even though the power statistics of large arrays of independent wind generators approach the normal distribution, as required by the Central Limit Theorem, they remain distinctly non-normal for small power levels near the hard lower bound at zero output. Dowds and co-authors note that many commercial wind integration studies for the United States assume normal distributions of low-wind events [7].

Some previous studies have characterized tail probabilities (i.e. the occurrence of low wind levels) empirically by examining historical wind-speed records. In a 1978 study of data from 25 weather stations in what was then West Germany, Molly found that the times during which total generation of arrays of hypothetical wind plants was zero declined from 1500–7200 hours per year for single sites to less than 5 h/y for arrays of 18 sites within the 800-km (N-to-S) national region [8]. A more recent study by Archer and Jacobson using wind-speed data from meteorological stations in the U.S. Midwest found that the incidence of average afternoon wind speed less than a typical turbine cut-in speed (i.e. $v < 3$ m/s) dropped from 7.6% of the time for single sites to 2.6% for three sites spread over a $120 \times 160$ km area, to 0% for eight sites spread over a $550 \times 700$ km area [9]. The duration curve they presented in follow-on work [10] indicates that wind generation was below 5% of turbine capacity 21% of the time for a single site, 10% of the time for a 7-site array, and 1.6% of the time for a 19-site array. In a study of the Nordic region using actual wind generation records, Holttinen found that while Denmark alone had production below 1% of capacity nearly 5% of the time during the years 2000–2002, the entire Nordic region never fell that low [11]. Using numerical-weather-model reanalysis data roughly corresponding to the territory of the Mid-continent independent system operator (MISO) Fisher *et al.* found that for a network of 108 sites the output level that could be counted upon all but 10% of the time was 7% of capacity during the winter and 3% of capacity during the summer [12]. These studies used historical data to characterize the tail probabilities, but don't offer a way to predict the effect of adding additional wind plants in new locations.

Recent investigations have focused on the effects of spreading arrays of wind generators over especially large distances. Kempton *et al.* [13, 14] considered an array of offshore wind plants distributed along the entire extent of the U.S. East Coast, while Fertig *et al.* [15] and Louie [16] evaluated the smoothing effect on wind generation of interconnections between independent system operators (ISOs) across the U.S. Huang *et al.* used reanalysis data to study the variability of coupled wind plants spread over the Great Plains of the U.S. from Montana to Texas [17]. A common feature of these studies is a sharp decline with increased geographic diversity of the fraction of time the aggregated wind power falls below small generation thresholds.

A number of studies have attempted to estimate the complete probability distributions of aggregated wind power, including the tails, in terms of parameters determined from the contributing generators. Justus and Mikhail [18] characterized an entire array of wind generators by a single "effective" array wind speed having a Weibull distribution with its shape parameter chosen to make its standard deviation $\sigma_N$ smaller than the standard deviation for a single site $\sigma_1$ according to the number of sites in the array and their average correlation $\bar{\rho}$: $\sigma_N = \sigma_1[(1 + \bar{\rho}(N-1))/N]^{1/2}$. They then modelled the array output power distribution by transforming the array wind speed distribution through a new power curve that cut in at a lower wind speed and reached rated capacity at a higher speed than did the turbines supposed to be deployed at the individual sites. This produces a complete model output-power probability density function; the "narrowing" of the wind-speed distribution and the "widening" of the power curve act in concert to greatly diminish the probability of low array power. Kahn used the extensive data and pioneering method of Justus and Mikhail to calculate the ELCC for wind arrays in California [3]. Sobolewski and Feijóo used non-parametric distributions defined by kernel estimators [19]. Carlin and Haslett pursed an alternate approach where they assumed the square-root of site wind speed was normally distributed, allowing the probability of zero array power to be calculated almost exactly from the characteristics and pair-correlations of contributing individual sites [20]. Hasche assumed array output-power could be described by the Pearson-family Type I (beta) distribution, and modeled the dependence of standard deviation, skewness, and kurtosis by using empirical functions to match the distance dependence pair and higher-order correlations [21]. The chosen beta distribution has the advantage that it naturally accommodates the bounds on array output power at zero and total turbine capacity, although tail probabilities were not explicitly investigated.

We demonstrate a novel method to quantify the tail of the distribution of aggregate wind power using Large Deviations Theory. This method provides a simple heuristic characterization of how the decline in occurrence of low-generation events depends on the number of sites being aggregated. Although our characterization in the present form is useful only for statistically-independent sites, it does not require making any assumptions about, or fitting to, the form of the probability distribution of wind speeds or wind power levels. In contrast, the methods proposed by Hasche [21] and by Carlin and Haslett [20] require Weibull distributions or Pearson-family Type I distributions, respectively. Our method allows us to more accurately characterize wind *power*, which is the convolution of a wind speed distribution and turbine power curve and not easily represented by common distributions. Carlin and Haslett noted "the effect of dispersal on the probabilities of zero and rated power is significantly more marked than on the coefficient of variation of windpower" a result we concur with and further quantify here. Our method explains this observation as a consequence of the differing dependence of tail and central behavior on the number $N$ of independent generators: exponential $N$-dependence of tail probabilities according to Large Deviation Theory in contrast to the $1/\sqrt{N}$ dependence of the coefficient of variation according to the Central Limit Theorem.

**Methods**

*Simulated wind-power data*

We investigate the number of hours per year that aggregate wind power is less than a chosen threshold by simulating the power output of arrays of widely separated wind plants using historic wind speed data. We select 9 wind sites in the continental United States, shown in Figure 1, according to three criteria: they have publically-available data from towers taller than typical meteorological stations (instrument height $h$ in Table 1), their mean wind speeds are similar (mean $\bar{v}$ in Table 1), and their simulated wind power outputs are poorly correlated (correlation coefficients in Table 1). We use 1-hour mean wind speeds for the period from January 2007 to December 2012, a total of $5.26 \times 10^4$ hours. Most sites have anemometers at multiple heights; we use the data from the height with mean wind speed closest to 6 m/s because that is the speed range for which data are available at the largest number of sites. We exclude any measurements identified as bad by the supplier, measurements less than 0 or greater than 40 m/s, or measurements inconsistent with those taken by other sensors on the same tower at the same time. This quality control excludes 9–38% of the data from the individual sites and leaves $1.46 \times 10^4$ hours when data are available from all sites simultaneously. The mean wind speeds from these nine sites, shown in Table 1, are lower than would typically be selected for commercial wind power development, but are the choices that provide the largest number of sites with similar wind speeds.

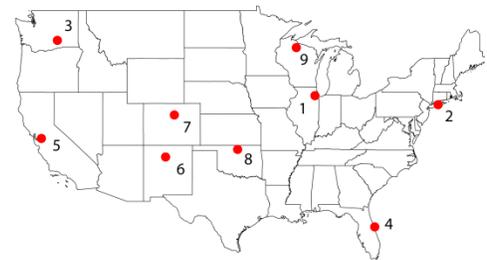

**Figure 1. Locations of nine tall-tower sites in the U.S. (See supplementary data for site details).**

| | site | $h$ (m) | $\bar{v}$ (m/s) | $\mu$ = cf | $\delta_0$ | $\delta_1$ | \multicolumn{8}{c}{correlation coefficients} |
| | | | | | | | 1 | 2 | 3 | 4 | 5 | 6 | 7 | 8 |
|---|---|---|---|---|---|---|---|---|---|---|---|---|---|---|
| 1 | Argonne | 60 | 5.4 | 0.27 | 0.11 | 0.02 | | | | | | | | |
| 2 | Brookhaven | 88 | 5.8 | 0.32 | 0.13 | 0.03 | 0.062 | | | | | | | |
| 3 | Hanford | 122 | 5.0 | 0.27 | 0.36 | 0.08 | −0.034 | −0.026 | | | | | | |
| 4 | Kennedy | 90 | 6.0 | 0.35 | 0.12 | 0.03 | 0.091 | 0.078 | 0.015 | | | | | |
| 5 | LLNL | 23 | 6.0 | 0.37 | 0.27 | 0.12 | −0.025 | −0.028 | −0.041 | −0.026 | | | | |
| 6 | Los Alamos | 92 | 4.6 | 0.23 | 0.38 | 0.04 | 0.049 | 0.002 | 0.077 | 0.008 | 0.004 | | | |
| 7 | NWTC | 80 | 4.8 | 0.22 | 0.37 | 0.06 | 0.070 | 0.048 | −0.077 | 0.052 | −0.070 | −0.007 | | |
| 8 | SGP | 25 | 6.1 | 0.36 | 0.15 | 0.08 | 0.170 | 0.031 | −0.014 | 0.071 | −0.043 | 0.240 | 0.055 | |
| 9 | WLEF | 122 | 6.3 | 0.41 | 0.13 | 0.03 | 0.220 | −0.010 | 0.029 | 0.029 | −0.055 | 0.065 | 0.023 | 0.061 |
| | "representative" | | | 0.31 | 0.23 | 0.06 | | | | | | | | |

Table 1. Data characteristics for each site: anemometer height $h$ above ground, average wind speed $\bar{v}$; for simulated wind power, the capacity factor cf, the fraction of time the turbine produces no power $\delta_0$, and the fraction of time it is at full power $\delta_1$; Pearson cross-correlation between sites. The parameters of the representative site are those of the histogram in Figure 2.

Wind power is simulated from the historical wind speed data using a turbine power curve based on the Vestas V110 2.0 MW turbine [22], with a cut-in wind speed of 3 m/s, reaching rated power normalized to 1 at wind speed of 11 m/s, and cut-out wind speed of 25 m/s. Details of the functional form of the power curve are given in the supplementary data. Hereinafter, we represent wind power as a fraction of total wind-generator capacity.

We count the number of hours per year that the simulated power aggregated from combinations of $N$ sites is less than a chosen threshold $p_0$, for $N$ ranging from 1 (individual site) to 9 (all sites combined), and plot $p_0$ vs. the count in the form of a generation duration curve, as in Figure 5. This curve, which depicts the same information as a cumulative distribution function (cdf), is a plot with the power threshold $p_0$ on the $y$-axis and the number of hours per year that aggregate power is less than $p_0$ on the $x$-axis. (Here we plot *hours less than threshold* rather than the conventional hours greater than threshold, but on an $x$-axis with its origin at the right and values increasing to the left. The curves thus retain precisely their conventional form and meaning but allow the use of a logarithmic axis to portray small duration values.) For each combination of sites, the duration curve was calculated by averaging the individual sites' simulated power (computed from the measured wind speed) at each hour (excluding any hour for which data was missing from one or more of the sites comprising that array).

To characterize a single site with behavior "representative" of the nine sites we also pooled all the hourly simulated power values from the nine individual sites (about $4.2 \times 10^5$ samples) and calculated a single histogram, as shown in Figure 2, taking care to separately accumulate those simulated power values of exactly 0 ($\delta_0$) or exactly 1 ($\delta_1$). We use a histogram with 72 bins (70 full-width and 2 zero-width for $\delta_0$ and $\delta_1$), but we show in the supplementary data that our results are not sensitive to the number of bins. For these nine sites the "representative" wind plant produces zero power approximately 2000 h/y, and full power about 520 h/y. Its capacity factor or average output power is $\mu = 31\%$. The 0.00–0.014 bin is empty because the power curve of the simulated turbine steps discontinuously from 0.0 to 0.014 at the cut-in wind speed.

**Large Deviations Theory model**

Although the simulated power outputs of the sites in Figure 1 are neither independent nor identical we nevertheless model the simulated aggregate wind power of an *N*-site array as the mean $\overline{P_N}$ of *N* independent identically distributed (i.i.d.) copies of a "representative" random variable *X* having the probability distribution given by the histogram shown in Figure 2. Large Deviations Theory (LDT) gives, under quite broad conditions, tight bounds on the probability that this mean is less than some small value (see Lewis and Russell [23] for an accessible introduction to LDT). According to LDT, $\Pr(\overline{P_N} < p_0)$ falls with *N* as $e^{-Q(p_0)N}$. The "rate function" $Q(p_0)$ is given as the Legendre transform of the random variables' cumulant generating function $\lambda(\theta)$:

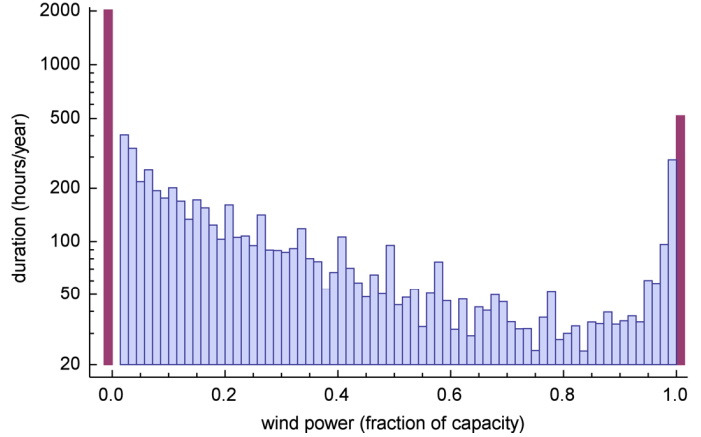

Figure 2. Histogram of simulated wind power for combined wind-speed records from all nine sites. Dark bars at each end represent instances of power being exactly 0 or exactly 1; bars are widened and offset for clarity. The distribution has mean $\mu$ = 0.31 and standard deviation $\sigma$ = 0.34.

$$Q(p_0) \equiv \sup_\theta [p_0 \theta - \lambda(\theta)] ; \quad \lambda(\theta) \equiv \ln\langle e^{\theta X}\rangle, \tag{1}$$

with $\langle \cdot \rangle$ denoting expectation. Although the exponential of the rate function captures the leading asymptotic dependence of probability on *N*, following Rozovsky [24] a more complete expression is given by:

$$\Pr(\bar{P} < p_0) = \frac{-1}{\vartheta \sigma(\vartheta)\sqrt{2\pi N}} e^{-Q(p_0)N}(1 + o(1)), \tag{2}$$

where $\vartheta$ is the value of $\theta$ at which the supremum in equation (1) is found, and the second derivative of the cumulant generating function gives $\sigma(\vartheta) = [\lambda''(\vartheta)]^{1/2}$. With bin heights $y_k$ for the histogram in Figure 2 normalized to represent the total fraction of samples in each of the 70 bins, we calculated the cumulant generating function as:

$$\lambda(\theta) \equiv \ln\left[\delta_0 + \delta_1 e^\theta + \sum_1^{70} y_k e^{(\frac{2k-1}{140})\theta}\right]. \tag{3}$$

Finding the maximum of equation (1) by numerical search gives *Q* and $\vartheta$; evaluating the second derivative of $\lambda(\theta)$ at $\vartheta$ lets us calculate the desired probability in equation (2).

In this simple model, the variability-reducing benefit of aggregating wind power from *N* sites is largely determined by the magnitude of rate function *Q*. As seen in Figure 3, *Q* rises as power threshold $p_0$ is decreased below mean $\mu$, indicating that modeled variability reduction through geographic diversity becomes more effective at smaller thresholds. The limiting value of *Q* at small thresholds can be understood by considering the probability that the output wind power of an *N*-site array is zero. This requires that its *N* sites are all simultaneously at zero power, which under our model's i.i.d. assumption has probability $(\delta_0)^N$, consistent with the calculated limit for $Q(0) = -\ln(\delta_0)$ resulting in $e^{-Q(0)N} = (\delta_0)^N$. Comparing the rate function calculated above with the rate function for a normal distribution having the same mean and variance shows the importance of the model's probability density being zero for negative wind power values. The normal-distribution rate function has a simple analytic expression [23] shown in Figure 3. The rate functions of both go to zero at $\mu$, but for thresholds ("deviations") away from the mean, the tail of the mean of normally-distributed variables, which extends to $-\infty$, declines at a slower rate than does the tail of the mean of our bounded wind-power variables. That is, for bounded distributions like wind power, the distribution's tails are thinner than those of the normal distribution. The effect of changes in the probability of zero-power output $\delta_0$ and in the output-power distribution on the rate function is further illustrated by the dashed curve, which shows the rate function calculated from the simulated wind power distribution for a site in Sweetwater, TX (see supplementary data) where higher average wind-speeds (7.9 m/s) result in the modeled turbine being below cut-in wind speed a much smaller 6.3% of the time compared to the $\delta_0$ = 23% for our "representative" site.

LDT provides good estimates of the probability that mean aggregate power $\overline{P_N}$ is less than some small value of $p_0$; for the purposes of LDT, $p_0$ is considered "small" if the probability that $\overline{P_N}$ is less than $p_0$ is also small, i.e. if $p_0$ is far from the centre of the distribution of $\overline{P_N}$. For the cases we investigate in this paper with $N \leq 9$ and the distribution shown in Figure 2 with $\mu = 0.31$, $p_0$ values less than 0.07 can be considered "small." On the other hand, when $p_0$ is close to the mean, the Central Limit Theorem tells us that, at least for large $N$, $\Pr(\overline{P_N} < p_0) \sim \Phi[(\mu - p_0)/(\sigma/\sqrt{N})]$, where $\Phi$ is the cdf of the unit normal distribution and $\mu$ and $\sigma$ are the mean and standard deviation of the distribution in Figure 2. We illustrate the ranges of $p_0$ for which LDT and a normal distribution provide good estimates of the probabilities with Figures S1 and S2 in the supplementary data.

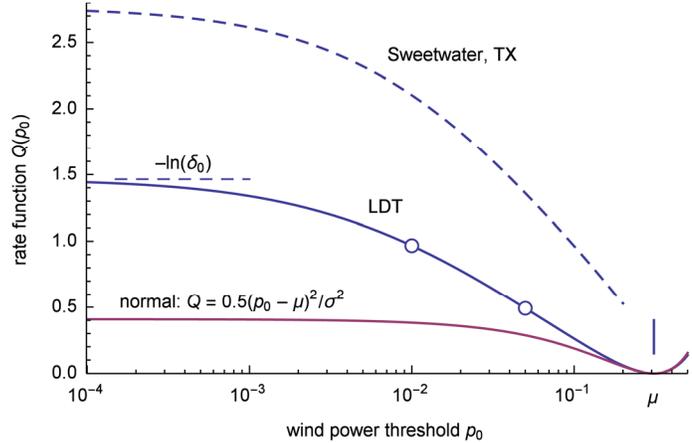

**Figure 3. Rate functions for the wind generation distribution of Figure 2 computed according to both LDT and the normal approximation. Symbols show the rate function values used for the LDT model curves in Figure 4. Also shown is the rate function for a site (Sweetwater, TX) with wind resource quality typical of a commercial wind plant.**

**Results and discussion**

The number of hours per year that simulated wind power aggregated from an $N$-site array is less than a given threshold $p_0$ decreases essentially exponentially with the number $N$ of aggregated i.i.d. sites, as shown in Figure 4. The box-plot shows the range of durations for all possible combinations of $N$ of the 9 sites shown in Figure 1, overlaid with theoretical curves computed under the assumption that the sites are uncorrelated and have identical distributions. The theoretical curves for thresholds $p_0$ of 0.01 and 0.05 are calculated using LDT with the rate-function values shown in Figure 3, because those thresholds are far from the mean power $\mu$. The theoretical curve for $p_0$ of 0.15, which is closer to the mean, is calculated using a normal distribution. Because we chose sites with low average wind speeds, the number of hours per year of low-wind events for our nine sites are significantly higher than for a typical commercial wind site, shown by the crosses. The correspondence of these results, calculated without the use of any adjustable parameters, to the simulated power data encouraged us to compare LDT results to the complete forms of the *N*-site generation duration curves.

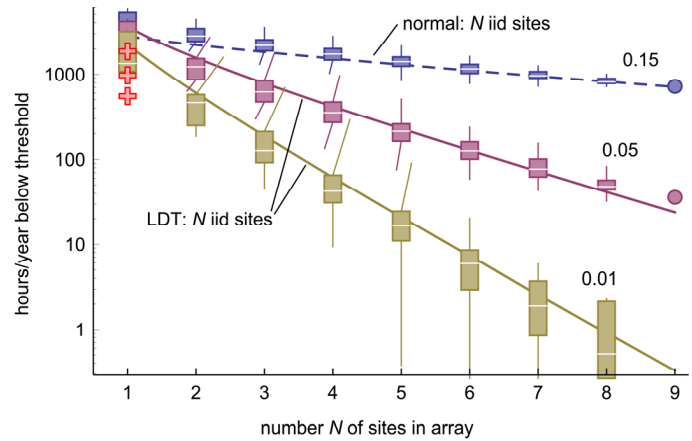

**Figure 4. Fraction of time wind power for an array of N sites is less than $p_0 = 15\%$, 5%, or 1% of total capacity. The number of low-power hours per year decreases approximately exponentially with $N$. Box and whisker symbols show the spread of sub-threshold hours of wind power simulated from measured wind speeds for various possible combinations of $N$ sites, with the whisker spanning minimum to maximum and the box the central two quartiles. Circles plot value for the unique 9-site combination. For $p_0 = 1\%$ the missing circle and cut-off whiskers and box indicate values of 0. Dashed and solid curves show theory assuming $N$ i.i.d. sites; dashed: normal distribution, $\Phi[(\mu - 0.15)/(\sigma/\sqrt{N})]$, solid: LDT, $1.69 e^{-0.49 N}/\sqrt{2\pi N}$, and $1.67 e^{-0.97 N}/\sqrt{2\pi N}$), respectively. Crosses plot sub-15%, 5%, and 1% durations for a site with mean wind speeds (7.9 m/s) more typical of a commercial wind power plant**

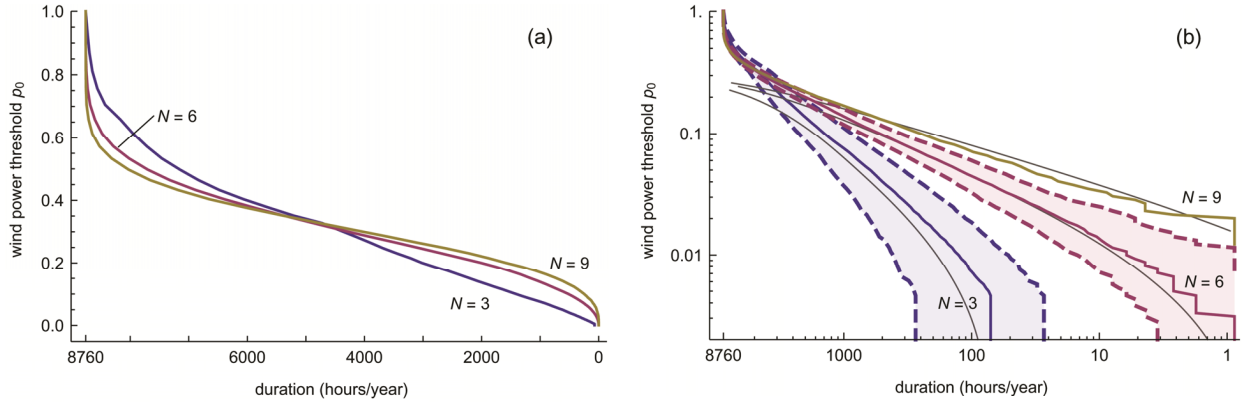

**Figure 5.** Generation duration curves for aggregate power from *N* sites (a) Linear axes. (b) Logarithmic axes: empirical (color) and LDT (assuming *N* i.i.d sites, black). Dashed lines and shading enclose 5–95% range of durations from different combinations of *N* sites; solid color line is the median of the empirical values.

Figure 5 plots generation duration curves for *N*-site arrays on linear (a) and logarithmic (b) axes. Each duration curve plots the fraction of hours (reversed *x*-axis) that the aggregate wind power is *less* than the value $p_0$ (*y*-axis). The linear plot (a), shows that this unconventional axis reversal nevertheless gives the conventional duration curve, as explained above in Methods. For both *N* = 3 and *N* = 6 there are 84 different combinations of sites possible. The colored lines in (b) plot the duration curves calculated from empirical data: solid lines show the median durations and dashed lines with shadings encompass the 5– 95% ranges. The thin black curves show the LDT model, again calculated assuming i.i.d. sites and without adjustable parameters from the representative distribution in the histogram of Figure 2. Horizontal "slices" of Figure 5 for chosen power thresholds ($p_0$ = 0.01, 0.05, 0.15) correspond to the curves shown above in Figure 4. According to the correlation coefficients listed in Table 1 our sites are not completely independent, with two of the site-pairs having correlation coefficients in excess of 0.2. Nevertheless, this small partial correlation seems not to prevent a close correspondence of the LDT curves to the median simulated wind-power data in Figures 4 and 5. Comparing the variances of the *N*-site arrays to the variances of the underlying individual sites also shows the partial correlation effects are modest. If the individual sites were uncorrelated the array variance should be equal to 1/*N* times the sum of the individual variances. The variance of the 9-site array simulated wind power (0.0154) is actually 7.6 times smaller than the average single-site variance of the distribution in Figure 2 (0.118), indicating that its behavior might be closer to that of an array of 8 uncorrelated sites.

**Conclusions**

The model presented here provides a quantitative basis for understanding the increase in firm capacity with geographic diversity. Our results demonstrate that aggregating wind plants can decrease the occurrence frequency of low-power events more dramatically than it decreases the magnitude of typical of variations around the mean. For weakly correlated sites we find the occurrence of low-power events in fact declines exponentially with *N*, in accord with Large Deviations Theory. For comparison, according to the Bienaymé Formula, the standard deviation of the mean decreases as $1/\sqrt{N}$. Thus, to decrease the probability of aggregated wind power falling below 1% of capacity by a factor of 20 for a 3-site array requires an increase in the number of aggregated sites from 3 to 6, at least for sites with characteristics similar to those investigated here. Cutting the standard deviation by a similar factor would require increasing the number of independent sites from 3 by a factor of 400, to 1200—almost certainly not possible.

This work may be useful in addressing planning for future modifications to the electric power grid. For our data, year-to-year variations in the simulated wind power distribution have only modest effects on the rate function (see supplementary data, Fig. S4). This may allow a grid planner to extrapolate the firm wind power capacity available with a given reliability from limited historical data. With regard to smoothing benefits in general, it is important to note the important caveat that our results do not calculate the time duration of individual low-wind-power events, i.e. they do not distinguish between ten one-hour periods and one ten-hour period of low power.

The empirical results presented here evidence good agreement with model results based on the assumption that the sites are independent. We presume that this agreement is a consequence of the weak correlation of our widely spread sites. However, relating the predictions of our model to the variability of real wind power plants depends critically on the extent to which the number of statistically independent sites is a good proxy for geographic diversity. The data we analyze inform speculation neither about the performance of arrays of more-closely spaced wind plants nor about achieving more than nine effectively independent sites within the contiguous U.S.

Additional work is needed to determine to what extent the methods and results presented above apply to situations with the higher inter-site correlations that typically result from clustering wind plants in areas with the best wind. Additional work is also needed to extend the methods and results we present to estimate the capacity value or ELCC of wind power, because they neglect the correlation of wind generation with electrical load.

**Acknowledgements**

This material is based upon work supported by the National Science Foundation under Grant No. 1332147 and by the Doris Duke Charitable Foundation, the Richard King Mellon Foundation, the Electric Power Research Institute, and the Heinz Endowments through the RenewElec project. The authors thank Prof. Julie Lundquist for directing them to many of the public sources of historical wind speed records used here. The authors also thank Argonne National Laboratory, Brookhaven National Laboratory, Lawrence Livermore National Laboratory, Los Alamos National Laboratory, National Renewable Energy Laboratory, the U.S. Department of Energy's Hanford Site, and its Atmospheric Research Measurement facility, NASA Kennedy Space Center, and the NOAA Earth System Research Laboratory WLEF site for providing wind speed data. They are acknowledged in more detail in the Supporting Information.

**Electronic Supplementary Information:**

# Is it always windy somewhere? Occurrence of low-wind-power events over large areas

Mark A. Handschy,*[a] Stephen Rose[b] and Jay Apt[b,c]

# Contents

## Large Deviations Theory vs. Central Limit Theorem

Figure S 1 below plots the empirical duration curve, the Large Deviations Theory (LDT) curve, and the normal distribution duration curve for $N = 6$ to illustrate how LDT provides a better model for the tail and the normal distribution provides a better model near the mean. In this case with $N = 6$ and a mean power of $\mu = 0.31$, the two models cross at a power threshold of $p_0 \approx 0.1$: LDT is a better model for lower thresholds and a normal distribution is a better model for higher thresholds. The normal distribution curve plots duration = $8760\Phi[(\mu - p_0)/\sigma_6]$. Figure S 2 plots the duration and power thresholds $p_0$ at which the LDT curve crosses the normal-distribution curve for various values of $N$.

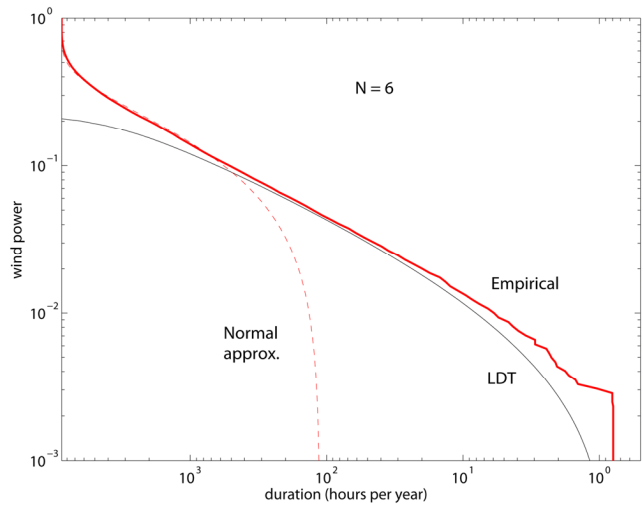

**Figure S 1.** Comparison of empirical generation duration curve for median simulated wind power from 6-site arrays with LDT model and with normal distribution of mean $\mu = 0.31$ and standard deviation $\sigma_6 = 0.34/\sqrt{6}$.

[a] *Enduring Energy, LLC, 5589 Arapahoe Ave., Suite 203, Boulder, CO 80303 USA.*
[b] *Department of Engineering and Public Policy, Carnegie Mellon University, 5000 Forbes Ave, Pittsburgh, Pennsylvania 15213 USA.*
[c] *Tepper School of Business, Carnegie Mellon University, 5000 Forbes Ave, Pittsburgh, Pennsylvania 15213 USA.*

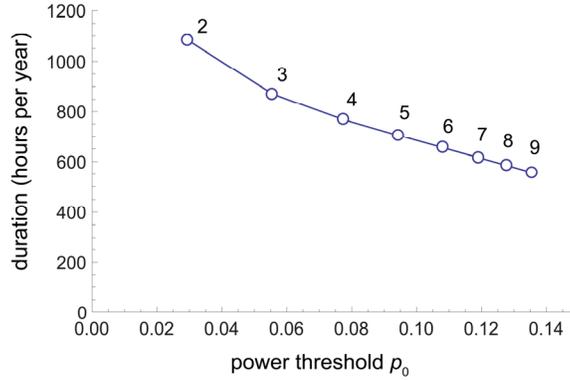

**Figure S 2.** Locus of cross-over between LDT model and normal distribution curves for various values of *N*.

## Sensitivity analysis for "representative" power distribution

In order to use LDT to model the distribution of the aggregate power of several non-identical wind plants, we developed a distribution for a hypothetical single site intended to be "representative" of all the individual sites. This "representative" distribution, shown in the main body of the paper as Figure 2, was calculated by pooling all the hourly simulated wind-power samples and then binning them into a single histogram. We show below that our results are not particularly sensitive to the data used to calculate the representative distribution.

Figure S 3 shows the results of LDT calculations using the power distribution of individual sites and using the representative distribution. It is clear that LDT using the "representative" power distribution matches the empirical duration curve (blue) far better than LDT using any single-site power distribution.

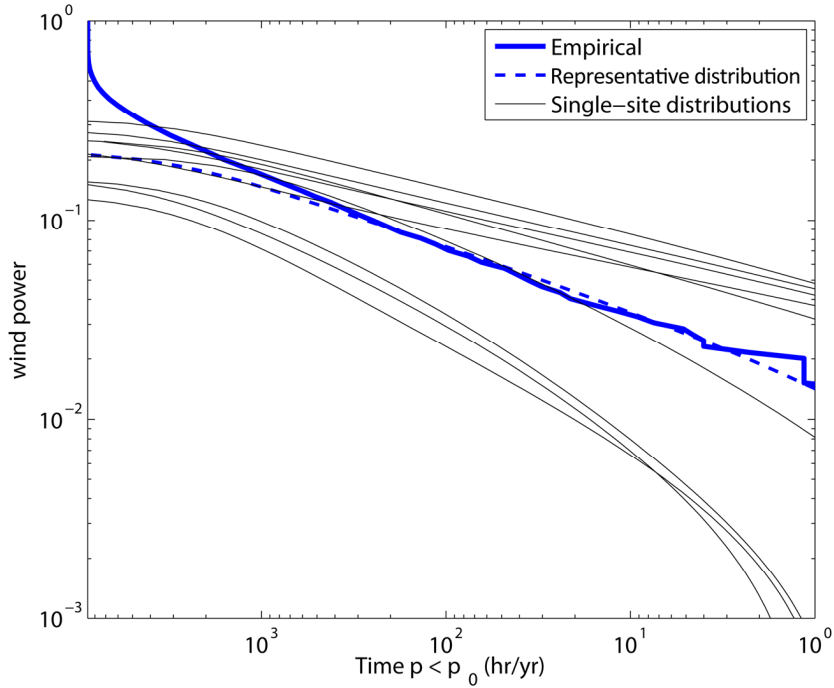

**Figure S 3.** LDT duration curves calculated using power distributions of each of the nine single sites along with the "representative" power distribution derived from all sites (Fig. 2 in main body), and 9-site aggregate empirical curve.

Figure S 4 shows the results of LDT calculations carried out as before, but instead of using the all-years "representative" distribution we substituted a histogram calculated from all sites but only for individual years. Each thin black line represents the LDT prediction from the pooled distribution for a different year. This figure shows that the distribution changes very little from year to year. This also suggests that a single year of data is probably sufficient to extrapolate the probabilities of events that occur less frequently than once per year.

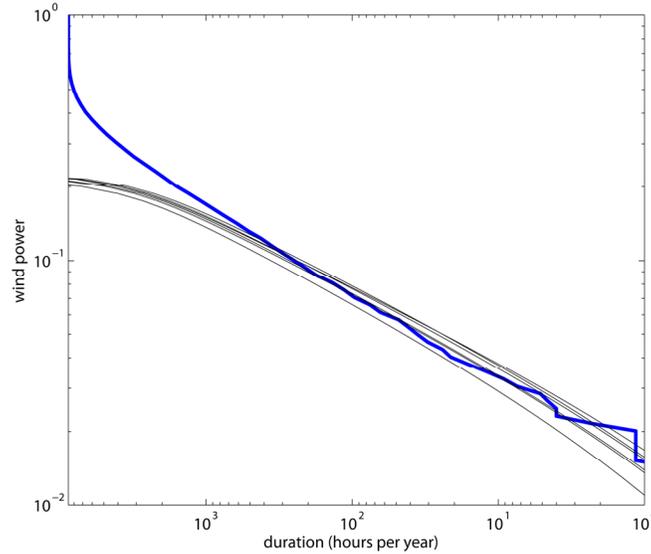

**Figure S 4. LDT duration curves calculated using power distributions derived from single years of the all-sites-combined data, and the 9-site aggregate (all-years) empirical curve.**

Figure S 5 shows the results of LDT calculations based on representative distributions calculated using different numbers of histogram bins (see histogram in Fig. 2 of the main body). The fit of the LDT duration curve to the empirical duration curve is not sensitive to the number of bins in the histogram.

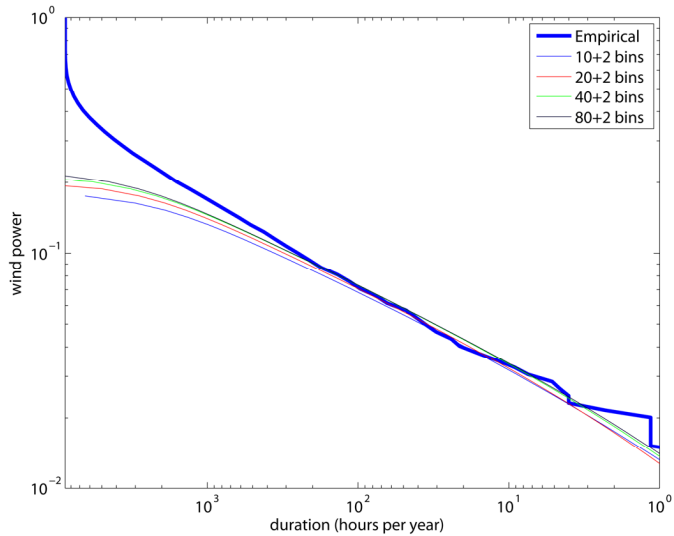

**Figure S 5. LDT duration curves calculated by pooling data from all nine sites, but using different bin widths than were used in Fig. 2 (main body), and 9-site aggregate empirical curve. Each histogram retains zero-width bins to accumulate incidence of output power levels of exactly zero and exactly one.**

## Turbine power curve

We transformed wind-speed samples to simulated wind power using the turbine power curve shown in Figure S 6, which approximates the manufacturer's curve for the Vestas V110 2.0 MW turbine. The turbine cuts in at $v = 3$ m/s, at which point its output steps discontinuously from 0 to 1.6% of capacity. For wind speeds greater than 11 m/s the output normalized by turbine capacity is unity, until the turbine cuts out at 25 m/s.

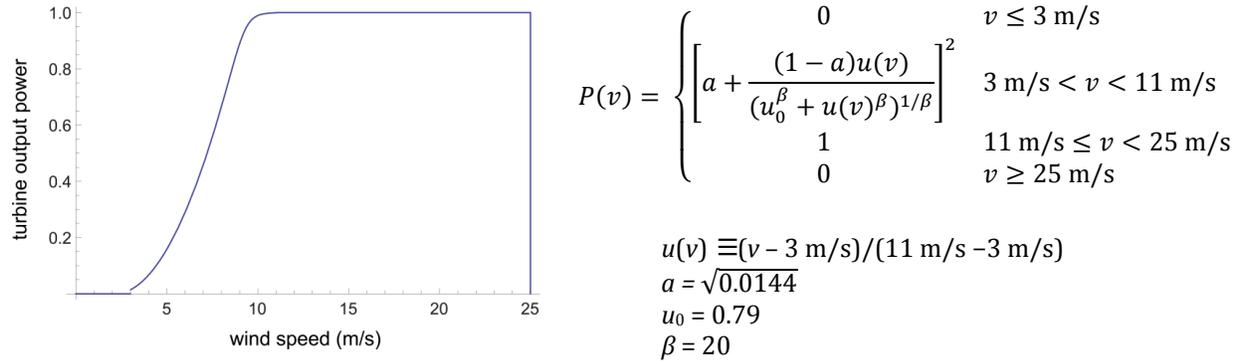

$$P(v) = \begin{cases} 0 & v \leq 3 \text{ m/s} \\ \left[a + \dfrac{(1-a)u(v)}{(u_0^\beta + u(v)^\beta)^{1/\beta}}\right]^2 & 3 \text{ m/s} < v < 11 \text{ m/s} \\ 1 & 11 \text{ m/s} \leq v < 25 \text{ m/s} \\ 0 & v \geq 25 \text{ m/s} \end{cases}$$

$u(v) \equiv (v - 3 \text{ m/s})/(11 \text{ m/s} - 3 \text{ m/s})$
$a = \sqrt{0.0144}$
$u_0 = 0.79$
$\beta = 20$

**Figure S 6. Model turbine power curve and parametric functional form used to transform historical wind-speed samples to simulated wind power.**

## Historical wind-speed data sites

**Table S 1. Wind-speed data-site locations and characteristics.**

| | | Location | Measurement height (m) | Mean wind speed (m/s) |
|---|---|---|---|---|
| 1 | Argonne National Laboratory | 41.702° N, 87.995° W | 60 | 5.4 |
| 2 | Brookhaven National Laboratory | 40.871° N, 72.889° W | 88 | 5.8 |
| 3 | ARM Southern Great Plains Central Facility | 36.606° N, 97.489° W | 25 | 5.0 |
| 4 | Lawrence Livermore Natl. Laboratory, Tower 300 | 37.675° N, 121.541° W | 23 | 6.0 |
| 5 | Hanford Site, Station 21 | 46.563° N, 119.600° W | 122 | 6.0 |
| 6 | Los Alamos National Laboratory | 35.861° N, 106.320° W | 92 | 4.6 |
| 7 | Kennedy Space Center Tower 313 | 28.626° N, 80.657° W | 90 | 4.8 |
| 8 | National Renewable Energy Laboratory(NWTC) | 39.810° N, 105.235° W | 80 | 6.1 |
| 9 | WLEF TV | 45.945° N, 90.273° W | 122 | 6.3 |
| | Sweetwater, TX, 51 Tall Tower South | 34.412° N, 99.646° W | 75 | 7.9 |

1. **Argonne:** Atmospheric and Climate Research Program, Environmental Science Division, Argonne National Laboratory. Downloaded on 2014 Jan 30 from:
   http://www.atmos.anl.gov/ANLMET/numeric/

2. **Brookhaven:** Personal communication with Scott Smith, Meteorological Services, Environmental and Climate Sciences Department, Brookhaven National Laboratory, 2014 Jan 30.

3. **Hanford:** Personal communication, Kenneth Burk, Hanford Site, Hanford Weather Station. 2014 Feb 18.